\documentclass[twocolumn]{aastex62}

\accepted{April 14, 2020}

\shorttitle{Submoons orbiting Kepler 1625b-I}
\shortauthors{Rosario-Franco et al.}

\begin{document}

\title{Orbital Stability of Exomoons and Submoons with Applications to Kepler 1625b-I}

\correspondingauthor{Marialis Rosario-Franco}
\email{mrosario@nrao.edu, marialis.rosario@mavs.uta.edu}

\author[0000-0003-0216-559X]{Marialis Rosario-Franco}
\affiliation {National Radio Astronomy Observatory, Socorro NM 87801, USA}
\affiliation{University of Texas at Arlington, Department of Physics, Arlington TX 76019, USA}  

\author[0000-0002-9644-8330]{Billy Quarles}
\affil{Center for Relativistic Astrophysics, School of Physics, 
Georgia Institute of Technology, Atlanta, GA 30332 USA}

\author[0000-0003-1975-9298]{Zdzislaw E. Musielak}
\affiliation{University of Texas at Arlington, Department of Physics, Arlington TX 76019, USA} 

\author{Manfred Cuntz}
\affiliation{University of Texas at Arlington, Department of Physics, Arlington TX 76019, USA} 
 


\begin{abstract}

An intriguing question in the context of dynamics arises: Could a moon possess a moon itself? Such a configuration does not exist in the Solar System, although this may be possible in theory.  \cite{Kollmeier2019} determined the critical size of a satellite necessary to host a long-lived sub-satellite, or submoon. However, the orbital constraints for these submoons to exist are still undetermined. \cite{Domingos2006} indicated that moons are stable out to a fraction of the host planet’s Hill radius $R_{H,p}$, which in turn depend on the eccentricity of its host's orbit. Motivated by this, we simulate systems of exomoons and submoons for $10^5$ planetary orbits, while considering many initial orbital phases to obtain the critical semimajor axis in terms of $R_{H,p}$ or the host satellite's Hill radius $R_{H,sat}$, respectively. We find that, assuming circular coplanar orbits, the stability limit for an exomoon is 0.40 $R_{H,p}$ and for a submoon is 0.33 $R_{H,sat}$. Additionally, we discuss the observational feasibility of detecting these sub-satellites through {photometric, radial velocity, or direct imaging} observations using the Neptune-sized exomoon candidate Kepler 1625b-I \citep{Teachey2018} {and identify how stability can shape the identification of future candidates}.

\end{abstract}



\section{Introduction}\label{sec:intro}

In stellar systems, planets orbit stars and moons orbit those planets in an hierarchical configuration (i.e., a satellite chain).  Naturally, one questions whether exmoons, or satellites, could host their own satellites (i.e., submoons) and thereby extend the chain one link further. In the solar system, only two planets (Mercury and Venus) are without moons, where there are $\sim$350 minor planets or asteroids that also have natural satellites\footnote{\url{https://minorplanetcenter.net/iau/mpc.html}}.  For the host bodies to hold onto their moons for long timescales, the satellites must reside within a restricted dynamical space (\citealt{Domingos2006}; hereafter DWY06), that depends on both the Roche and Hill radius as natural boundaries.  The ideal conditions for moons occur when the tidal stresses of the surrounding larger objects are small as to not affect the orbital stability of the satellite. Tidal stress can make or break a satellite by altering its spin and orbital parameters \citep{Reid1973}, as well as potentially disrupting the satellite completely. Tidal evolution in the context of star-planet-moon systems has been studied in the general case of moons orbiting exoplanets \citep{Barnes2002,Sasaki2012,Sasaki2014,Piro2018}.  Depending on the configuration, moons may migrate inward towards the Roche radius to break-up or collide with their host planet, migrate outward towards the Hill radius leading to expulsion from the system, or they might migrate only temporarily \citep{Barnes2002,Sasaki2012,Piro2018}.  The tidal migration of the moons is also tied to the host planet's rotation, which in turn is affected by stellar tidal friction \citep{Ward1973,Burns1973}.

Submoons (i.e, satellites of satellites) appear to be absent within the solar system, but may exist elsewhere in the Milky Way.  To detect such objects, a good starting place will be in the search for exoplanets with moons (i.e., exomoons) and leverage the strengths of the current proposed techniques for discovering exomoons.  Such methods include microlensing \citep{Han2002,Han2008,Liebig2010}, direct imaging \citep{Cabrera2007,Agol2015}, cyclotron radio emission \citep{Noyola2014,Noyola2016}, photometric transit timing \citep{Sartoretti1999,Kipping2009a,Kipping2009b}, and radial velocity \citep{Van2018}.  The most promising of these methods are through photometric transit timing and cyclotron radio emission; however, either a degeneracy exists within these methods that makes the detection ambiguous or the current state of technology has yet to achieve the necessary precision to obtain an adequate observation.

Currently, there is only one exomoon candidate, Kepler 1625b-I \citep{Teachey2018}, whose host planet is Jupiter-like belonging to a system with a Sun-like host star.  The initial vetting of this candidate used data obtained from the Kepler mission as well as follow-up observations with the Wide Field Camera (WFC) of the Hubble Space Telescope.  \cite{Teachey2018} examined several different models after the data was corrected for known systematics of the respective instruments aboard the telescopes and concluded that a Neptune-sized satellite represented the best-fit model quite well.  Other observers have contested these results suggesting that the exomoon signal is an artifact of the analysis \citep{Kreidberg2019} or possibly a misidentified event from other exoplanets orbiting the same host star \citep{Heller2018}.  Upon inspection of all the data, \cite{Heller2019} found strong statistical evidence using the Bayesian Information Criterion which favored the planet-moon model over the single-planet model, thus suggesting the existence of an exomoon.  

While \cite{Hamers2018} identified tidal capture as an alternate formation pathway for which the Jovian planet could have acquired such a large moon instead forming the satellite from a circumplanetary disk \citep{Canup2006}.  {Through numerical simulations, the wide orbit of Kepler 1625b-I could be reproduced if the exomoon candidate begins as a circumstellar co-orbital body with the original core of the giant planet Kepler-1625b.  The smaller co-orbital body is later pulled down and captured into a circumplanetary orbit, becoming an exomoon \citep{2019Hansen}.}  Even the habitability of a putative Earth-like exomoon was studied \citep{Williams1997,Forgan2018,Forgan2019} given that the Jupiter-like host exists within the habitable zone of its host star.  Evidently, additional observations are needed to confirm the exomoon signal but this may also prove difficult depending on the observed geometry of the exomoon's orbit relative to the host star's radius \citep{Martin2019}.

There are challenges for observing exomoons and likely additional hurdles for submoons, but the current exoplanet candidate Kepler 1625b-I provides the means to address potential outcomes numerically.  DWY06 produced a fitting formula to identify the outer boundary for stability concerning exomoons that was informed by earlier works examining the stability of similar architectures \citep{HW1999}.  In the past decade, the potential stability of exomoons was probed for individual systems \citep{Quarles2012,Cuntz2013} and more general studies \citep{Quarles2020} similar to those by \citeauthor{HW1999}.  {The stability of artificial submoons (e.g., satellites orbiting Earth's moon) is also dependent on the mass distribution within the host satellite, where mass concentrations, or mascons, can act to perturb the orbit of the submoon and limit its potential stability, especially for low-altitude orbits.  The Lunar Prospector Mission hinted at these effects \citep{2001Kono} and the GRAIL mission showed prominent mascons, the largest of which are associated with nearside basins, as well as the broad structure of the highlands.  It is likely that orbits of submoons orbiting terrestrial satellites can be unstable due to perturbations the host planet, as well as, perturbations created by mascons in the host satellite. For the purpose of our studies, we do not consider the effect of mascons, where we assume host planets and moons exert a uniform gravitational field.}

In this work,  we address whether this formalism can be improved for exomoons or expanded to include submoons.  Besides the outer stability limit, \cite{Kollmeier2019} showed that not just any submoon would be orbitally stable due to tidal interactions, where we investigate the long-term stability of submoons.  Future space and ground based telescopes could overcome the observing some challenges for the photometric, radial velocity, and/or cyclotron radio emission methods, where we identify the thresholds that such methods should meet in order to confirm the possible existence of a submoon.

In Section \ref{sec:methods}, we outline our numerical methods that include N-body simulations as well as the possible effects on observations.  We discuss our results for the stability of exomoons in Section \ref{sec:stab_sat} and submoons in Section \ref{sec:stab_sub}.  The possible limits due to tidal migration for a exomoon and submoon orbiting Kepler 1625b are explored in Section \ref{sec:tides}.  Section \ref{sec:obs} explores the potential observational scenarios for Kepler 1625 assuming that a stable exomoon and submoon exist there.  Finally, We conclude this work with a summary along with the prospects for future observations in Section \ref{sec:summary}.

\section{Methods} \label{sec:methods}
We examine the stability of submoons through N-body simulations for both short ($10^5$ yr) and long (10 Myr) timescales.  In order to investigate the short timescale, we use the IAS15 integrator in \texttt{REBOUND} \citep{Rein2012,Rein2015}, which uses an adaptive time-step to ensure adequate numerical accuracy within our simulations.  For evaluating the orbital evolution on long timescales, we use a modified version of the N-body integrator \texttt{mercury6} \citep{Chambers2002} that has been designed to efficiently evolve a similar hierarchy.  In \texttt{mercury6}, we adjust the parameters for the switch-over function so that the evolution of the submoon is evaluated via the Bulirsch-Stoer (BS) algorithm and the orbits of the other bodies (planet and moon) are evolved through the Wisdom-Holman (WH) portion of the hybrid algorithm.  Since stable submoons must orbit their parent bodies closely, this algorithm is well-suited for this particular hierarchy.  From this division, we choose an initial timestep of 5\% of the exomoon orbital period such that the WH integration also maintains a high accuracy over long timescales.

This setup also assumes that some formation pathway exists for both the exomoon and submoon (e.g., tidal capture; \citet{Hamers2018}), where the submoon can persist past the early stages of formation of the host planet; even though such events are probably rare.  \citet{Sasaki2012} showed that a Jupiter-mass planet can host an Earth-mass satellite and such a satellite will be safe from tidal effects on 10 Gyr timescales at planet separations larger than 0.3 AU.   We follow the results from \citet{Kollmeier2019} and prescribe an exomoon radius such that a long-lived submoon will be negligibly affected by the tidal interactions with the host satellite using the following:
\begin{equation} \label{eqn:KR19}
    R_{sat} \geq \left[\frac{39 M_{sub} k_{2,sat} T \sqrt{G}}{2(4\pi\rho_{sat}/3)^{8/3} Q_{sat}} \left( \frac{3M_{p}}{(f a_{sat})^3} \right)^{13/6}
    \right]^{1/3},
\end{equation}
which includes the exomoon radius $R_{sat}$, submoon mass $M_{sub}$, planet mass $M_p$, exomoon love number $k_{2,sat}$, exomoon density $\rho_{sat}$, exomoon tidal quality factor $Q_{sat}$, exomoon semimajor axis $a_{sat}$, and system lifetime $T$.  Following the system parameters from \cite{Teachey2018}, a Jovian (4 M$_J$) planet orbits a Sun-like star (1.04 M$_\odot$) on a $\sim$287 day ($\sim$0.86 AU) orbit.  At such a large separation, the impact of tides from the host star is small on satellites (and submoons) of Kepler 1625b.  For the exomoon candidate, Kepler 1625b-I, we consider a 22 day (or $a_{sat}$ = 0.024 AU) orbit around the Jovian host and a Neptune-like mass ($m_{sat}$ = 0.0564 M$_J$) for the exomoon, consistent with the observations from \citet{Teachey2018}.  From the calculated radius in Equation \ref{eqn:KR19}, most submoons would likely not support a substantial atmosphere so we use an asteroid-like density (3 g/cm$^3$) to calculate the mass of the submoon, which is very small ($m_{sub}$ = 1 $\times$ $10^{-6}$ M$_\oplus$) compared to the host planet and exomoon.

Our general simulations for exomoons use our short timescale ($10^5$ yr) in a similar manner as DWY06, where a Jupiter-mass planet begins on a Earth-sized orbit ($a_p = 1$ AU) and hosts an Earth-mass exomoon.  The orbital eccentricity of the planet and exomoon are varied from 0.0 to 0.5 in steps of 0.01.  Both orbits begin coplanar to one another ($i_p=i_{sat}=0^\circ$) with the argument of pericenter and ascending node orbital elements set to zero ($\omega=\Omega=0^\circ$).  The planet begins with its mean anomaly equal to zero ($MA_p=0^\circ$), while twenty values  of the exomoon's mean anomaly are randomly selected from a uniform distribution from $0^\circ$ to 360$^\circ$.  The simulations that differ in orbital phase are used to calculate the fractional stability $f_{stab}$ of an initial condition, where a simulation is given either zero weight or full weight depending on the numerical outcome (unstable or stable, respectively).  We use the planet's Hill radius ($R_{H,p}=a_p\left(\mu_p/3\right)^{1/3}$; where $\mu_p=M_p/M_\odot$) to scale our exomoon results.  Using the prograde results from DWY06, we increment the exomoon semimajor axis $a_{sat}$ from 0.25 $R_{H,p}$ to 0.55 $R_{H,p}$.  A simulation is terminated and marked unstable, if the parent-satellite separation $r$ is less than the parent radius (i.e., collision) or greater than the Hill radius of the parent body (i.e., escape).

\begin{figure}
    \centering
    \includegraphics[width=\linewidth]{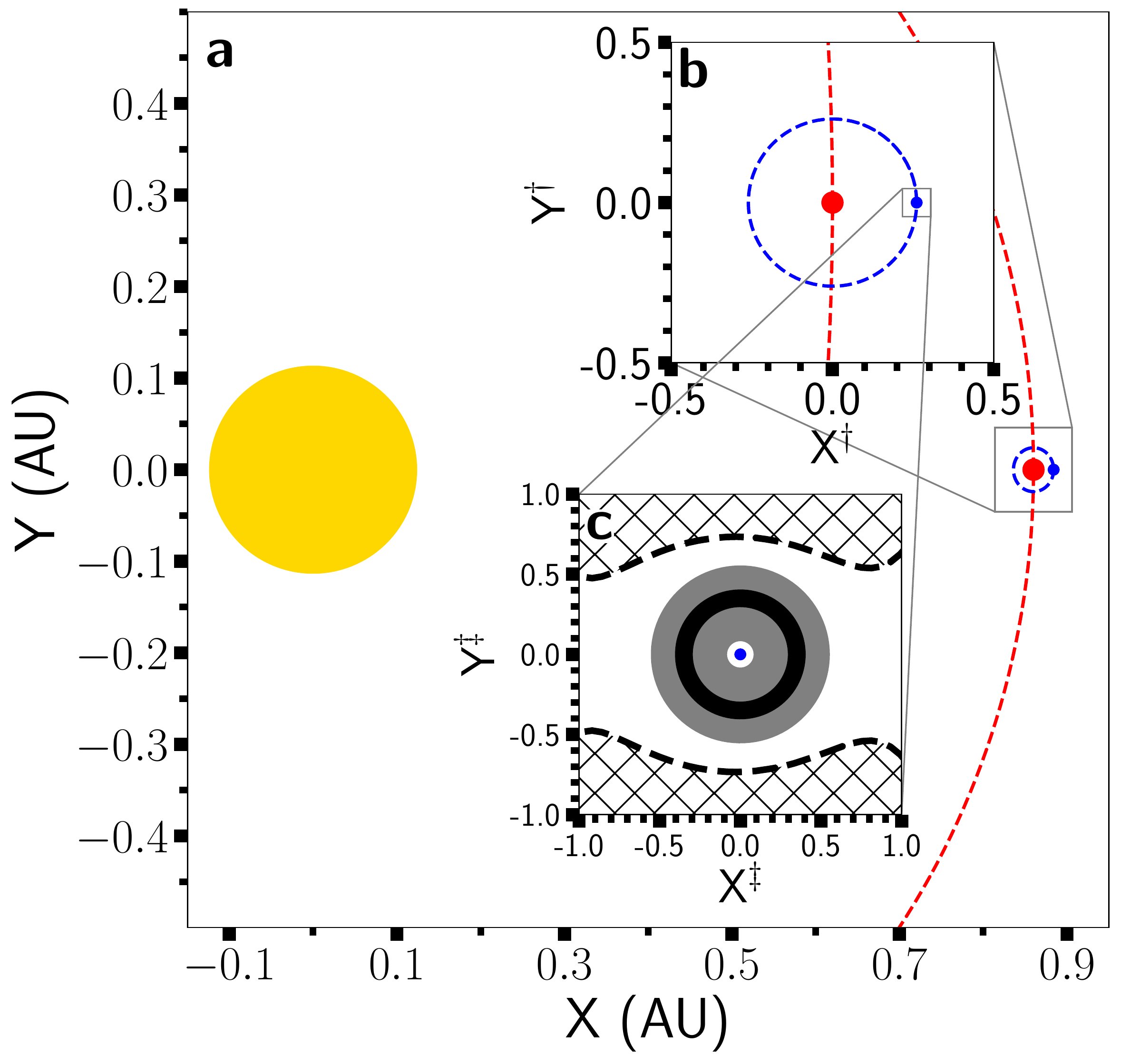}
    \caption{Diagram of our initial setup for circular, coplanar orbits: (a) Kepler 1625b (red dot) orbits Kepler 1625 (yellow dot), (b) an exomoon (blue) orbits Kepler 1625b and (c) a submoon orbits the exomoon within either the gray or black annulus in our short- (100 kyr) or long-term (10 Myr) simulations, respectively. The black dashed curves (in panel c) mark the zero velocity boundary, which represents the maximum theoretical extent of submoons, and the hatched region denotes the forbidden zone.  The dagger ($\dag$) and double dagger ($\ddag$) symbol denote that the distance units are in terms of the planet's Hill radius $R_{H,p}$ and exomoon's Hill radius $R_{H,sat}$, respectively. Note that the size of the orbits are to scale, while the size of the dots are not.} 
    \label{fig:overplot2}
\end{figure}

We perform two sets of simulations for submoons using a Neptune-like exomoon host, where one set is a grid search similar to the method describe above and the other explores more general initial orbits (i.e., all angles drawn from distributions).  For the first set, we select $a_{sat}$ for a 22 day orbit (i.e., 0.22 $R_{H,p}$) and vary the submoon semimajor axis $a_{sub}$ from 0.05 $R_{H,sat}$ to 0.55 $R_{H,sat}$ using the exomoon's Hill radius ($R_{H,sat}=a_{sat}\left(\mu_{sat}/3\right)^{1/3}$; where $\mu_{sat}=M_{sat}/M_p$).  We choose 0.05 $R_{H,sat}$ as the lower limit for $a_{sub}$ because it is approximately the Roche limit, assuming a Neptune-like and asteroid-like density for the host exomoon and submoon, respectively.  Our second set explores a more limited range for $a_{sub}$ ranging from 0.3 $R_{H,sat}$ to 0.4 $R_{H,sat}$, uses the system parameters for Kepler 1625b-I, and draws all angles from distributions.  \cite{Teachey2018} suggested that the exomoon's orbit could be inclined relative to the orbital plane of the planet and thus, we draw the eccentricities and mutual orbital inclinations from Rayleigh distributions, whilst the rest of the non-fixed orbital parameters are drawn from uniform distributions.  Figure \ref{fig:overplot2} illustrates the extent of our initial setup, where Fig. \ref{fig:overplot2}c shows the putative boundaries within the context of the circular restricted three body problem \citep{Eberle2008,Musielak2014}.

\section{Results \& Discussion} \label{sec:results}
\subsection{Stability Limits of Exomoons and Submoons} \label{sec:stab}

An aspect that can be studied without the need of a complete theory of satellite formation is the orbital evolution of satellites since the survivors of such evolution can be potentially observed.  In this context, one of the goals of this work entails determining the stability boundary for orbits of possible exomoons and submoons around the exomoon candidate Kepler 1625b-I.  The stability boundary is a useful tool because it places a constraint on the maximum orbital period allowed for (sub)satellites, while tidal interactions can limit the minimum orbital period.
DWY06 derived expressions of the critical semimajor axis ($a_{E}$) in units of the planet's Hill radius $R_{H,p}$, where satellites located beyond the critical semimajor axis would escape the gravitational influence of the host planet.  More specifically, their work produced an expression for stable prograde and retrograde orbits, where the eccentricities of the planet $e_{p}$, and the satellite $e_{sat}$ are included \citep[see Equation 5 and 6 from][]{Domingos2006}.   Retrograde orbits that are typically associated with the irregular satellites of Jupiter \citep{Jewitt2007} also have high eccentricities.  This is a supremely limiting factor for the stability of exomoons and submoons given the hierarchical nature of our problem and thus retrograde orbits are not within the scope of this work. 

\subsubsection{Exomoon Stability Limits Revisited} \label{sec:stab_sat}
To understand the stability of submoons, first we must address the stability of their host exomoons.  Thus, we perform a suite of simulations varying the initial orbital phase $MA_{sat}$ and the semimajor axis $a_{sat}$ of the exomoon, as well as the eccentricity $e_p$ of the host planet.  We define the quantity $f_{stab}$ as the fraction of twenty simulations with randomly chosen $MA_{sat}$ that survive for $10^5$ yr \citep[i.e.,][]{Quarles2018,Quarles2020}.  Figure \ref{fig:stab}a show the results of our simulations, where $f_{stab}$ is color-coded and the white cells mark when $f_{stab}=0$.  The largest $a_{sat}$ (with $f_{stab}=1$) for a given $e_p$ marks the stability boundary and is followed by a transition region (colored cells).  The black dashed curve marks the best fit solution from DWY06 for prograde exomoons and the red solid curve represents our best fit solution to the stability boundary using the Levenberg-–Marquardt algorithm in the \texttt{curve fit} function within the \texttt{scipy} module.  The light gray curves mark the uncertainty in our estimation.  The first two rows of Table \ref{tab:form_stab} provide quantitatively the coefficients for each curve using ${a_{crit}} (R_{H,p}) = c_1(1-c_2e_p)$ as the model.  Our results agree with DWY06 when considering a single choice for the orbital phase of the exomoon, but widely disagree when we approach the stability more probablistically using $f_{stab}$.  The first coefficient $c_1\approx0.4$ (i.e., stability limit for circular orbits) differs by $\sim$20\%, while the second coefficient $c_2$ for the $e_p$ term differs by only $\sim$10\%.  These differences with DWY06 are beyond the uncertainties given and decreases their estimates for the maximum exomoon mass to only 30\% of the reported values (DWY06; see their Table 1).  \cite{Payne2013} also found $c_1\approx0.4$ when investigating exomoons within dynamically packed planetary systems, but their results are ambiguous between choaticity and stability through the use of the Lyapnuov exponent.    

Figure \ref{fig:stab}b demonstrates our results when the exomoon's eccentricity $e_{sat}$ also varies and allows us to define a new model ${a_{crit}} (R_{H,p}) = c_1(1-c_2e_p -c_3e_{sat})$ similar to DWY06.  The color-coded cells mark the largest $a_{sat}$ (with $f_{stab}=1$) for a given set of initial eccentricities $e_{sat}$ and $e_p$.  As one might expect, the dependence on $e_p$ is stronger than for $e_{sat}$.  This is shown quantitatively in third row of Table \ref{tab:form_stab}, where $c_2/c_3 \approx 6$.  Additionally, we find that the second and third rows agree (within the uncertainties), which provides a sanity check for our numerical results.  We provide a fitting formula that better matches more robust numerical simulations for exomoons, but such a formula is reliable in characterizing a population of exomoons and can be inaccurate when describing particular systems.  As a result, we suggest a different approach using a lookup table or interpolation map that relies less on statistical averaging, which occurs in deriving a single fitting formula.


\subsubsection{Stability Limits for Submoons} \label{sec:stab_sub}
\cite{Kollmeier2019} previously studied the possible tidal evolution and stability of submoons following the analytical treatment for the Solar System moons \citep{Murray1999} and exomoons \citep{Barnes2002,Sasaki2012}.  However, part of their analysis depends on the stability limit for submoons, where small changes in $f$ (Eqn. \ref{eqn:KR19}) are amplified by the large exponent.  We acknowledge that high uncertainties remain in other terms (e.g., $k_{2,sat}$ or $Q_{sat}$), but changes in those values result in a much smaller change to $R_{sat}$ due to the smaller exponent.

To determine the stability limit, we follow a similar methodology as our investigation of exomoon stability (see Section \ref{sec:stab_sat}).  The major differences are that the submoon semimajor axis is incremented in units of the host satellite's Hill radius $R_{H,sat}$, the submoon's initial orbital phase $MA_{sub}$ is chosen randomly while the host satellite's mean anomaly is not ($MA_{sat}=0^\circ$), the host satellite's mass is set to nearly equal to Neptune ($M_{sat} \approx 18M_\oplus$) and the host satellite's semimajor axis is no longer varied ($a_{sat} \approx 0.015$ AU).  Figure \ref{fig:stab}c demonstrates the stability of a submoon, when only the planetary eccentricity and submoon semimajor axis are varied (identical color-code as Fig. \ref{fig:stab}a).  Highly stable initial conditions (black cells) cluster first near $a_{sub} \sim R_{H,sat}/3$ for low planetary eccentricity ($e_p\lesssim0.2$) and then more widely at $a_{sub} \sim R_{H,sat}/4$.  Due to the hierarchical nature of these systems, overlap of mean motion resonances can destabilize submoons in a manner similar to planets in binary systems \citep{Mudryk2006,Quarles2020}.  Hence a gap forms when the period ratio of the satellite to the submoon equals an integer ($P_{sat}/P_{sub} = \sqrt{3/f^3}$) and $f$ is a fraction of the host satellite's Hill radius.  The fourth row of Table \ref{tab:form_stab} shows the results of a linear fit, where the first coefficient $c_1 = R_{H,sat}/3$ is the most meaningful.  

Figure \ref{fig:stab}d demonstrates how the stability limit varies with changes in both the planetary and satellite eccentricity.  The orange region for low eccentricity corresponds to conditions where the stability island near $a_{sub} \sim R_{H,sat}/3$ can persist, where most choices of eccentricity push the stability limit to lower values (green to blue region).  Eccentricity values near 0.5 (gray region in Fig. \ref{fig:stab}d) are wholly unstable because the stability limit lies within the Roche limit of the host satellite.  The final row of Table \ref{tab:form_stab} provides coefficients for a fitting function, but such a model employs some averaging.  Specifically, the fitting function removes the plateau at low eccentricity (orange region) and smooths out the transition from 0.25 $R_{H,sat}$ to 0.20 $R_{H,sat}$ at high eccentricity of the satellite (light-green to blue region) in Fig. \ref{fig:stab}.  Comparing the fourth and last row in Table \ref{tab:form_stab}, there is a significant difference and the same sanity check that we employ for exomoons does not hold.  The difference arises because the host satellite's eccentricity is now allowed to vary and it plays a significant role in the stability of its submoons.  By default, we would expect that $c_2$ in the exomoon case should map onto $c_3$ for submoon stability.  This is mostly true, where some dilution occurs due to the main plateau at low eccentricities and the host planet still has some influence on the submoon.

\cite{Kollmeier2019} assumed that the stability limit for submoons to largely match that of exomoons when scaled by the appropriate Hill radius.  This is clearly not the case, where their results underestimate the true limits in the host satellite's radius.  Additionally, our Equation \ref{eqn:KR19} differs from \citeauthor{Kollmeier2019} by a factor of $1/3$ in the denominator that was {due to  a typographical error in the manuscript (S. Raymond 2020, private communication).  If we use our stability limit for $f$} (assuming nearly circular orbits), the minimum satellite radii reported are increased by a factor $k$ ${(=(0.4895/0.33)^{13/2} \sim 13}$).  Relaxing the assumption for nearly circular orbits to allow for eccentric orbits for the host planet and satellite, the denominator of the correction factor $k$ can be smaller, which makes the importance of the stability limit even greater.

\subsubsection{Long-term Stability of Submoons}
Our study of the stability of submoons is inherently limited by computational resources, where some may want to see billion year simulations to decide whether an initial condition is long-term stable.  However, a different approach is to use an analytic formula to ensure stability against tidal migration \citep[i.e.,][]{Barnes2002} and perform N-body simulations for roughly a billion orbits of the submoon.  Submoons that survive this timescale are likely to be extant in the face of either tides or gravitational perturbations.

We focus on region between 0.3 $R_{H,sat}$ to 0.4 $R_{H,sat}$ of the host satellite and perform simulations for 10 Myr, or 2 billion orbits of the submoon.  The remaining initial conditions for the submoon are chosen randomly, where $e_{sub}\lesssim 0.2$ and the inclination relative to the host satellite's orbital plane is $\lesssim$15$^\circ$ (see Section \ref{sec:methods}).  The initial semimajor axis for the host planet and satellite follow parameters from Kepler 1625b, but the remaining orbital elements are chosen randomly.  We evolved $\sim$1,500 initial conditions for this longer simulation timescale and find that the stable submoons cluster around $a_{sub} \approx 0.33 R_{H,sat}$ with low eccentricity $e_{sub}\lesssim 0.05$.  The inclination differences between the orbital planes of the planet, satellite, and submoon have a negligible effect on the long-term stability when they are significantly below $40^\circ$.

\subsection{Parameter Limits for Exomoons and Submoons in Kepler 1625} \label{sec:tides}
The tidal migration of moons depends on the assumed value for the tidal Love number $k_2$ and the tidal quality factor $Q$ of the host body.  The tidal Love numbers for Solar System bodies have largely been estimated \citep{Lainey2016} so that relatively accurate analogs can be used for our purposes.  However, the tidal quality factor $Q$ is very uncertain and thus, we are forced to vary this parameter over a huge range ($10^2-10^5$).  Despite this uncertainty, we can identify constraints for the exomoon candidate Kepler 1625b-I and explore the limiting mass of a putative submoon.

\begin{figure}
    \centering
    \includegraphics[width=\linewidth]{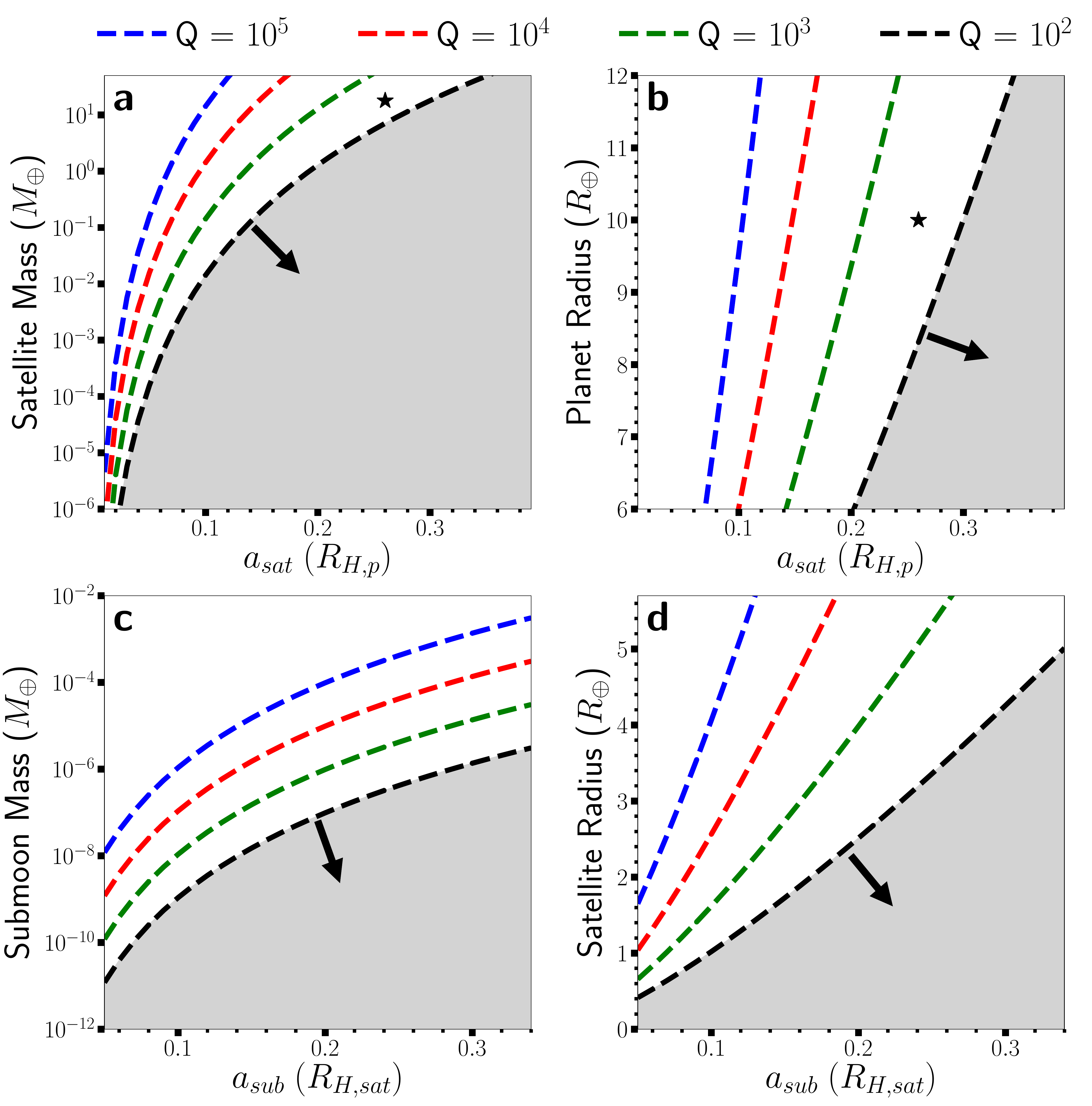}
    \caption{Limits on the parameters for exomoons (a and b) and submoons (c and d) orbiting Kepler 1625b using constraints due to tides \citep{Barnes2002} and orbital stability.  The dashed curves mark the maximum value as a function of either the planetary Hill radius $R_{H,p}$ or the satellite Hill radius $R_{H,sat}$.  The color-code for the curves denote the assumed tidal quality factor $Q$ of the respective host body, while the gray region shows a stable region for most reasonable assumptions for $Q$ and the arrows indicate the direction of increasing stability.  The tidal love number for the planet is assumed to be Jupiter-like ($k_{2,p}\approx0.54$; \cite{Lainey2016}) and the satellite is assumed to be Neptune-like ($k_{2,sat}\approx0.12$; \cite{Gavrilov1977}).  The black star symbols mark the estimated system parameters for the exomoon candidate Kepler 1625b-I \citep{Teachey2018}.  }
    \label{fig:limits}
\end{figure}

Figure \ref{fig:limits} shows the upper limits for the (a) exomoon mass, (b) host planet radius, (c) submoon mass, and (d) exomoon radius up to respective stability limit for coplanar circular orbits over a 9 Gyr timescale.  Each set of curves uses the system parameters from \cite{Teachey2018} to define the properties of the host body and also indicates the change in the upper limit with an assumed value for $Q$ (color-coded).  Figs. \ref{fig:limits}a and \ref{fig:limits}b have a star symbol that denote the location of Kepler 1625b and its candidate moon within the parameter space.  If we assume that Kepler 1625b is Jupiter-like in its tidal quality, then values below (or to the right of) the red dashed curve in Figs. \ref{fig:limits}a and \ref{fig:limits}b are allowed.  The exomoon parameters from \cite{Teachey2018} suggest that the candidate exomoon is similar to Neptune and thus values below the green dashed curves in Figs. \ref{fig:limits}c and \ref{fig:limits}d are allowed.  The maximum submoon mass is $3 \times 10^{-5} M_\oplus$, which is slightly less massive than the asteroid Vesta.  The minimum exomoon and submoon separations are 0.14 $R_{H,p}$ and 0.2 $R_{H,sat}$, respectively, given the measured radius of each host body and reasonable value of $Q$ (red or green dashed curve, respectively).

\subsection{Detection Thresholds for Submoons}\label{sec:obs}
 Although \cite{Kollmeier2019} introduced a formalism for the stability of a planet-moon-submoon, the possibility of observing such sub-satellites has not been addressed. The observation and confirmation of exomoons has proved challenging through current proposed techniques, as mentioned in Section \ref{sec:intro}. The exploration of exomoon and submoon detection is necessary to contextualize our work within existing observational capabilities. In this section, we use current exomoon detection techniques and determine the noise thresholds necessary for a detection of a submoon. In particular, we use the transit and radial velocity (RV) methods, which have been prolific in exoplanet confirmations. Additionally we discuss the challenges that submoon detection in radio wavelengths poses. 
 
 \subsubsection{Transit Method}
 To explore the feasibility of detecting submoons through transits we utilize Python packages \texttt{REBOUND} and \texttt{batman}. We use a sample from our long-term (10 Myr) integrations to produce the synthetic light curves (see Table \ref{tab:syn_IC}). The package \texttt{batman} \citep{2015Kreid} computes transit light curves for a single exoplanet using analytic expressions introduced by \citet{2002A&M} and include quadratic or nonlinear limb darkening.  We use the relative locations of each body on the sky plane in our rebound simulations to determine how much area covers the stellar disk and approximate the net flux (exoplanet + exomoon) using a single planet with an equivalent area (i.e., homeomorphic area).  We account for partial eclipses, or blends, by finding the intersectional area between two disks on the sky (e.g., exoplanet + star or exoplanet + exomoon) to correct the equivalent area for possible double counting.  

\begin{figure}
    \centering
    \includegraphics[width=\linewidth]{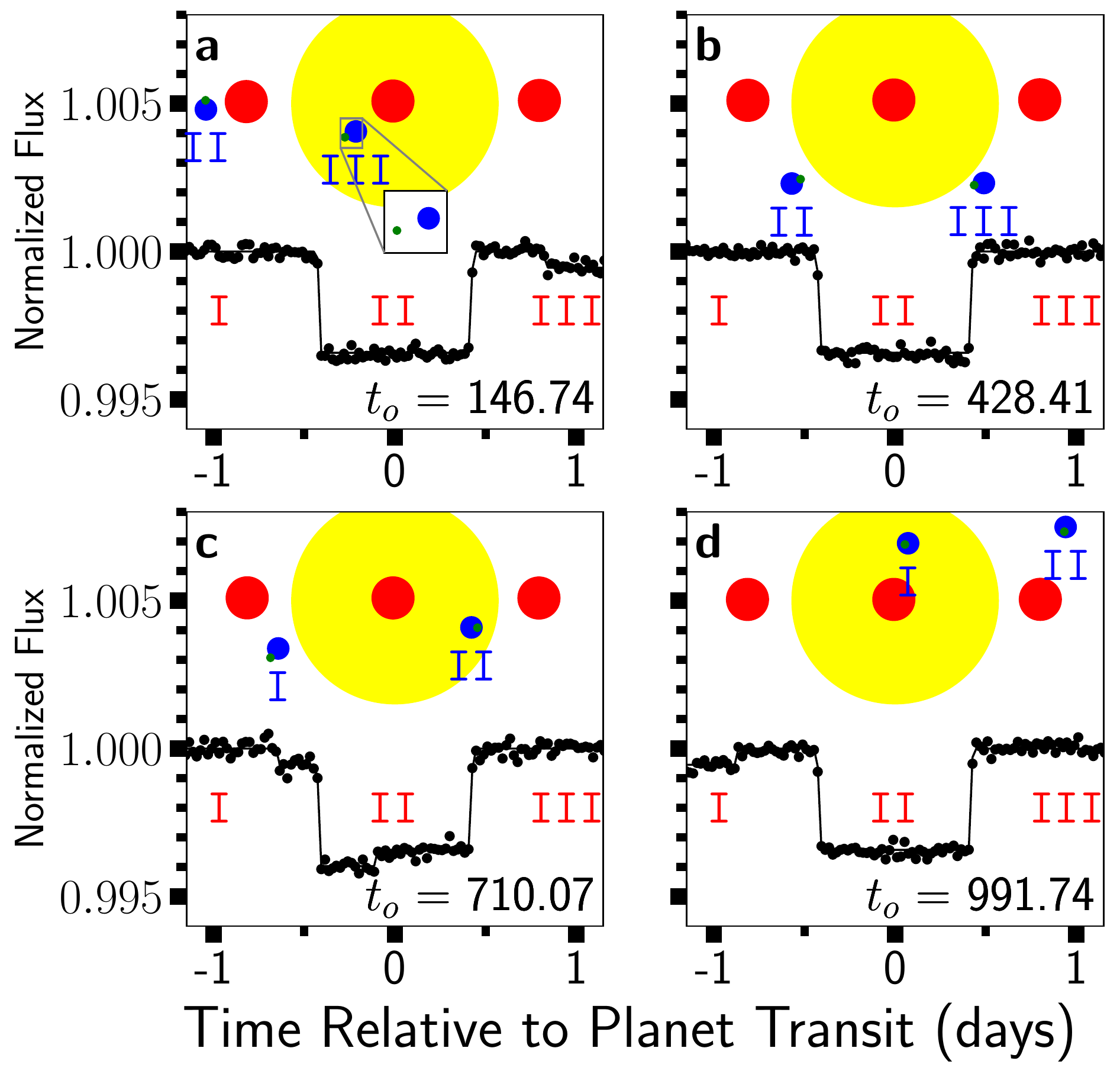}
    \caption{Simulated lightcurves of an exomoon system similar to Kepler 1625b-I at four different epochs $t_o$ (a-d).  Each panel illustrates the position of the exoplanet (red), exomoon (blue), and submoon (green; see zoomed inset) on the sky, where the location of the exomoon is annotated (with {\texttt{I}, \texttt{II}, or \texttt{III}}) in blue that corresponds to time epochs in red.  As a result, panels (a) and (d) show exomoon transits in addition to planetary transits.  Panel (c) demonstrates a blended transit, where the exomoon moves faster on the sky than the host exoplanet.  The transits of submoons are negligible given the current limits on photometric precision.  }
    \label{fig:light_curve}
\end{figure}

We simulate the lightcurves over a four year period (similar to the Kepler mission), identify four planetary transits around the host star, and evaluate the contribution of the satellites (exomoon and submoon) to the transit depth.  The synthetic light curves in Figure \ref{fig:light_curve} show satellite transits in addition to planetary transits, where Figs. \ref{fig:light_curve}a and \ref{fig:light_curve}d illustrate when the satellites transit $\pm1$ day relative to the planetary transit $t_o$.  

The illustrations in Figure \ref{fig:light_curve} are annotated {(\texttt{I}, \texttt{II}, or \texttt{III})} to identify the relative location of the satellites.  Due to stability constraints on the submoon (see Section \ref{sec:stab_sub}), the exomoon-submoon separation is always less than a stellar radius ($a_{sub} \approx R_\star/8$; $R_\star = 1.73R_\odot$) which implies that submoons, in general, will always transit as long as their host satellite transits too.  The 125 km submoon produces a transit depth of 0.01 ppm and is much smaller than the expected photometric noise in 30 min cadence observations.  Figure \ref{fig:light_curve}b shows only a planetary transit because the exomoon separation and relative inclination are large enough for transits to not occur at every planetary transit.  Figure \ref{fig:light_curve}c illustrates the difference in velocity on the sky plane between the planet and satellites, where the satellites transit, followed by the planet + satellites, and ending only with the planet.

\subsubsection{Radial Velocity Method}
We study the effect of satellites (exomoon and submoon) on potential radial velocity (RV) measurements of Kepler 1625 using \texttt{REBOUND} to produce synthetic radial velocities that consider the reflex motion of either the star or the planet.  Traditional RV measurements quantify the reflex motion of a star due to a planet (see \cite{2010F&D} for a review), but \cite{Van2018} recently showed that RV measurements taken in a direct imaging campaign could, in principle, detect the reflex motion of an exoplanet due to an exomoon.  We perform four separate numerical simulations over a single planetary period: (1) star-planet, (2) planet-satellite, (3) satellite-submoon, and (4) planet-satellite-submoon so that we can identify the reflex motion using the line-of-sight velocity $v_z$ of a body directly.

\begin{figure}
    \centering
    \includegraphics[width=\linewidth]{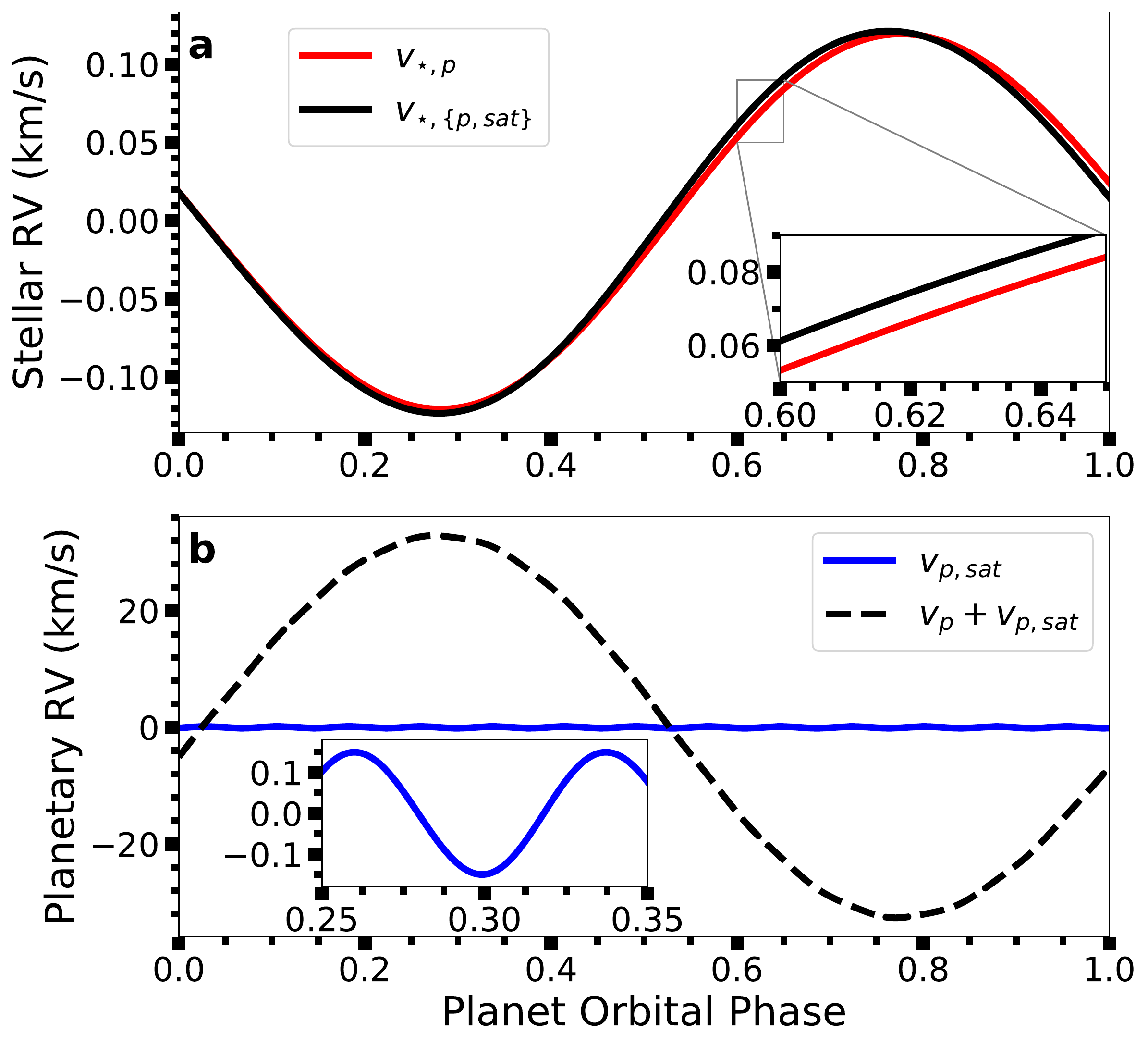}
    \caption{Synthetic radial velocity (RV) curves over a single planetary orbit showing the (a) reflex motion of the star and the (b) reflex motion of the planet.  The curves in panel (a) mark the RV of the star solely due to the planet (red), the RV including the indirect perturbations from the exomoon (black), and the inset panel highlights the $\sim$8 m/s difference between the curves.  Panel (b) demonstrates the RV signature from possible direct imaging observations (black dashed) following \cite{Van2018}, where the satellite induces $\sim$150 m/s of variation (blue) relative to the planetary center-of-mass velocity $v_p$.  The reflex velocity of the exomoon due to perturbation from a 125 km submoon is extremely small, $\sim$0.3 mm/s, and likely beyond the capabilities of next-generation instruments.}
    \label{fig:RV}
\end{figure}

The stellar reflex motion due to a planet $v_{\star,p}$ in Figure \ref{fig:RV}a shows an RV semiamplitude is $\sim$120 m/s (red curve), where the inclusion of a large satellite $v_{\star,\{p,sat}\}$ (black) produces similar results.  The exomoon produces a slight shift ($\sim$8 m/s) in the RV curve indirectly by causing the planet to accelerate coherently with its motion induced by the host star.  Assuming optimistic observational conditions, this difference in RV is detectable using current-generation spectrographs.  The impact of a submoon on the stellar reflex motion is negligible because the submoon causes a $\sim$0.3 mm/s oscillation in the host satellite's velocity.

As \cite{Van2018} showed, direct imaging observations could uncover exomoons by obtaining RV measurements of the host exoplanet.  Figure \ref{fig:RV}b demonstrates the expected RV curve from such measurements, where the planetary reflex velocity $v_{p,sat}$ (blue) has a semiamplitude ($\sim$150 m/s) slightly larger the stellar RV from the host planet, $v_{\star,p}$.  This is possible because the mass ratio for the planet-satellite is larger than for the star-planet and the exomoon is much closer to its host body.  However, RV measurements of the exoplanet directly will also include the center of mass velocity $v_p$ added to the planetary reflex velocity (i.e., $v_p+v_{p,sat}$) and is represented through the black dashed curve in Fig. \ref{fig:RV}b.  Isolating the exomoon signal would require $\sim$0.5\% precision in the RV measurements, which \cite{Van2018} shows is possible with an 8-m class telescope, high resolution spectrograph, and modern adaptive optics systems.  Our submoon produces a negligible effect on the planetary RV, where even the most massive submoons allowed by tidal migration would still produce miniscule shifts in the RV curve and is too small to be identified with current and future instruments. 

Although our work arrives at the same general conclusion as \cite{Van2018}, we do note that the most extreme RV semiamplitudes presented by \citeauthor{Van2018} are not likely realized because tidal migration (see Fig. \ref{fig:limits}a) over the lifetime of the system would have caused the exomoon to: (1) collapse onto the host planet, (2) migrate outward past the stability boundary, or (3) migrate outward to produce smaller RV semiamplitudes.

\subsubsection{Radio Emission}
Another detection method takes advantage of the radio signal that an an exomoon like the candidate Kepler 1625b-I would emit due to the  interaction of a satellite with its host planet's magnetic field. Following \citet{Noyola2014}, we calculate intrinsic power $P_s$ and incident flux $S$ of such a radio emission for the exomoon candidate orbiting Kepler 1625b. 

The expected power for Kepler 1625b-I will depend on several parameters, such as the magnetic field strength $B_s$, the plasma speed $V_0$, and the plasma density $\rho_s$ where each depends on the separation of the exomoon from its host planet.  \cite{Noyola2014} showed the detectability of Io-like exomoons orbiting a Jupiter-analog, where we can augment their calculation \citep[][see their Equation 2]{Noyola2014} through appropriate scale factors.  Io tightly orbits Jupiter at only 0.008 $R_{H,p}$, where the intrinsic power $P_s$ is 5 GW.  Signals from such a system with $S\approx 50$ $\mu$Jy are detectable at 50 MHz and only 15 light-years away.  Kepler 1625b-I is more widely separated (0.26 $R_{H,p}$) from its host planet and is much larger (4 $R_\oplus$), which yields a power of 3.2 GW assuming an Io-like plasma density, which is slightly less than Io.  The power would rise to 150 GW, if the exomoon orbits at the inner limit (0.12 $R_{H,p}$) set by tidal migration given a Jupiter-like dissipation (see Fig. \ref{fig:limits}a).  However the incident flux is also inversely proportional to the distance $d$ squared to the system (i.e., $S\propto P_s/d^2$) and Kepler 1625 is approximately 8000 light-years away.  As a result, the highest expected flux from the Kepler 1625 system is $10^{-4}$ times smaller than a Jupiter-Io analog, or $\sim$5 nJy.  Applying our values in Table \ref{tab:syn_IC}, the estimated flux drops even lower to 10 pJy due to a much weaker magnetic field at 0.26 $R_{H,p}$.

We find that regardless of the intrinsic power $P_s$ of the planet-moon interaction, its distance from Earth poses the biggest challenge to obtain a detection.  We refrain from exploring the effect of a submoon in the radio emission of the moon, as its radius is too small (125km) to contribute to the intrinsic power.  Even if the submooon could boost the power by an order-of-magnitude through plasma sharing between satellites \citep{Noyola2016}, the incident flux would still remain incredibly small due to the large distance to the system.

\section{Summary \& Conclusions} \label{sec:summary}
The orbital stability of exomoons (satellites) and submoons (sub-satellites) largely depends on the orbital eccentricity of the respective host body and is limited to a fraction of the host body's Hill radius, $R_H$.  {Although, a larger dependence on the (sub-)satellite's eccentricity may arise in multi-body systems (e.g., Galilean moons).}  We perform N-body simulations for $10^5$ planetary orbits and use twenty random initial orbital phases to determine a stability limit (in $R_H$), or the largest, stable semimajor axis, of exomoon and submoon systems.  As a result, the stability limit for an exomoon is 40\% of the host planet's Hill radius ($R_{H,p}$) and for a submoon is 33\% of the host satellite's Hill radius ($R_{H,sat}$), assuming circular, coplanar orbits.  Additionally, we determine constraints on the physical properties of exomoons or submoons so that they are extant over a wide range of tidal migration scenarios.  Exomoons are detectable given current photometric capabilities (e.g., Kepler or $TESS$), where large submoons ($\sim$400 km) require less than 0.3 ppm photometric precision for a Sun-like star.  Both exomoons and submoons will require next-generation radial velocity facilities, where the RV amplitude on the host star for exomoons is small and for submoons is negligible.  The flux through radio emission through interactions of Kepler 1625b-I on the planet's magnetic field would be incredibly small, [$10^{-9}-10^{-12}$] Jy, and would require extremely long integration times ($>10^3$ hr) with a telescope like the Square Kilometer Array. However, if the system were closer ($\sim$4.6 pc) the emission would be detectable since the incident flux would be much larger and would also require significantly lower integration time.

\begin{figure}
    \centering
    \includegraphics[width=\linewidth]{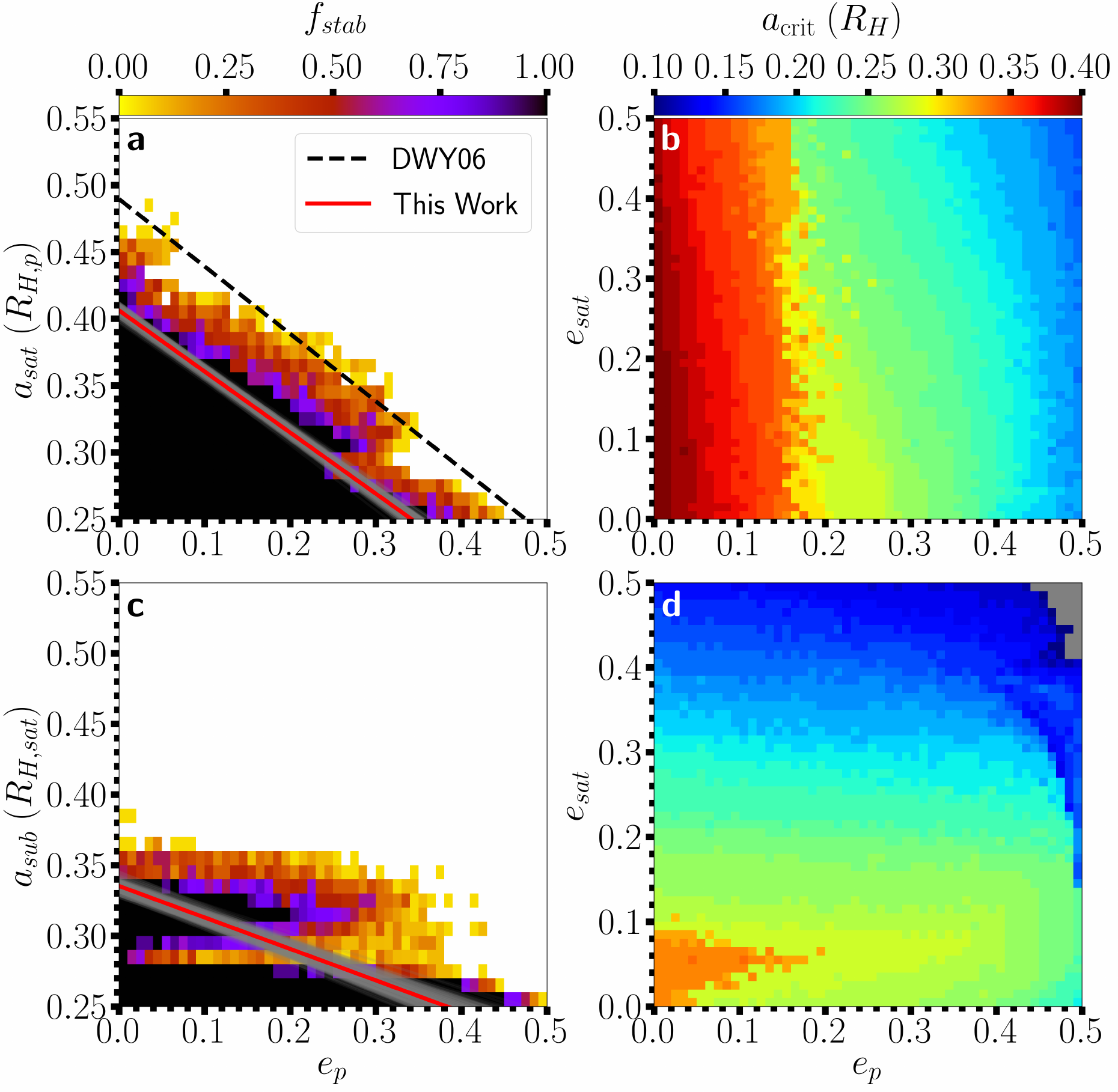}
    \caption{Stability boundaries using $a_{sat}$ for exomoons (a) and $a_{sub}$ for submoons (c) scaled by their respective Hill radius $R_H$. The color-code $f_{stab}$ in panels a and c denotes the fraction of initial conditions that are stable for $10^5$ yr from our simulations, where the white cells mark when $f_{stab}=0$.  The ${a_{crit}}$ values (color-code) in panels b and d represents the largest semimajor axis with $f_{stab}=1$ for a given pair of initial eccentricity values ($e_p$ and $e_{sat}$). The black dashed curve shows the empirically determined stability boundary by \cite{Domingos2006}.  The red solid curve denotes best fit for the stability boundary from our simulations, where the gray solid curves illustrate our uncertainty (see Table \ref{tab:form_stab} for values).}
    \label{fig:stab}
\end{figure}

DWY06 performed the quintessential study of exomoon stability.  However they probed only a single trajectory (orbital phase) for a given pair of initial conditions ($a_{sat},e_{sat}$) and caused them to overestimate the outer boundary ${a_{crit}}$ ($a_E$ {in their notation}) of stability by 20\%.  Our simulations are evolved for 10$\times$ more planetary orbits and consider twenty trajectories per initial condition (see Fig. \ref{fig:stab}\footnote{{The data for this figure are available on\dataset[GitHub]{https://github.com/Multiversario/satstab}, along with an example Python script that illustrates how to make a lookup function using our data.}}).  Using these simulations, we update the coefficients for the fitting formula used by DWY06 (see Table \ref{tab:form_stab}).  In addition, we expand upon the methodology to consider the stability limit for submoons (in $R_{H,sat}$) and find that submoons are stable to only $R_{H,sat}/3$ (see Table \ref{tab:form_stab}). 

Analytical tidal migration estimates from \cite{Barnes2002} depend on prior estimates of the stability limit as a fraction $f$ of the host planet's Hill radius and are typically used to identify the limiting physical characteristics of exomoons (DWY06).  These expressions raise $f$ to a very high exponent (6.5), where large deviations can result from small changes.  The major difference for exomoons is that large satellite masses with ($0.4R_{H,p}<a_{sat}\leq0.49 R_{H,p}$) are not viable due to tidal migration.  \cite{Kollmeier2019} based their analysis for submoons on \cite{Barnes2002} and DWY06, where the differences for submoons can be much larger due to a less smooth stability surface (see Fig. \ref{fig:stab}).  We use Equation 8 from \cite{Barnes2002} and the exomoon candidate system Kepler 1625b-I \citep{Teachey2018} to determine the upper mass limit (70\% the mass of Vesta) and upper radius limit (375 km) for submoons.  We also show that a Neptune-like exomoon would be orbitally stable around Kepler 1625b for most reasonable assumptions on the tidal quality ($Q\gtrsim$10$^3$) of the host planet.

We evaluate the detection prospects of exomoons and submoons orbiting Kepler 1625b through a range of methods, such as those using photometry, spectroscopy, and radio emission.  In the photometric method, the transit depth of the exomoon candidate (Kepler 1625b-I) is distinguishable from the host planet transits, but a submoon would require an extreme photometric precision below 1 ppm.  We produce synthetic RVs that show the stellar reflex motion or the planetary reflex motion because the latter could be measured using large telescopes with state-of-the-art direct imaging capabilities.  Through both of these measures, we find that the presence of exomoons is identifiable, where an 8 m/s difference arises in the stellar RV and 150 m/s oscillation in the planetary RV.  Recent observations have uncovered star-planet interactions through radio emissions \citep{Vedantham2020}, but such techniques require a much higher sensitivity (nJy scale) to make a detection of an exomoon orbiting Kepler 1625b. 

{It is worth noting that these observational techniques carry their own biases and in addition to stability, these biases may affect the discovery of exomoons and submoons.  The transit method is biased towards well-separated large (sub-)satellites, the RV method has a bias towards more massive bodies, and direct imaging includes both of these biases.  Our study for exomoons and submoons shows that orbital stability favors these biases and it makes sense that the first exomoon candidate by \cite{Teachey2018}, Kepler 1625b-I, is large and massive.  However, these features might not be common in the general population of exomoons or submoons.} 

\acknowledgments
{The authors thank Sean Raymond for his thorough review with comments that enhanced the quality of the manuscript.  We are grateful to Jason Barnes for useful comments and discussion.}  M.R.F acknowledges support from the NRAO Gr\"{o}te Reber Fellowship and the Louis Stokes Alliance for Minority Participation Bridge Program at the University of Texas at Arlington.  This research was supported in part through research cyberinfrastructure
resources and services provided by the Partnership for an Advanced Computing Environment (PACE) at the Georgia Institute of Technology.

\software{mercury6 \cite{Chambers2002}, batman \cite{2015Kreid}, rebound \cite{Rein2012,Rein2015}}

%








\bibliographystyle{aasjournal}
\bibliography{references.bib}

\begin{deluxetable}{lccc}
\tablecaption{Coefficients for the Critical Semimajor Axis  \label{tab:form_stab}}
\tablehead{\colhead{} &  \colhead{$c_1 \pm \sigma_1$} & \colhead{$c_2  \pm \sigma_2$} & \colhead{$c_3 \pm \sigma_3$} } 
\startdata
Exomoon (DWY06$^\star$) & $0.4895\pm0.0363$ & $1.0305\pm0.0612$ & $0.2738\pm0.0240$  \\
Exomoon & $0.4061\pm0.0028$ & $1.1257\pm0.0273$ &  -- \\
Exomoon & $0.4031\pm0.0007$ & $1.1230\pm0.0041$ &  $0.1862\pm0.0050$ \\
\hline
Submoon &  $0.3352\pm0.0036$ & $0.6642\pm0.0360$ &  -- \\
Submoon &  $0.3210\pm0.0008$ & $0.2757\pm0.0060$ &  $1.0687\pm0.0050$ \\
\enddata
\tablecomments{{The coefficients ($c_1 -c_3$) and uncertainties ($\sigma_1 -\sigma_3$)} from $^\star$\cite{Domingos2006} are listed using the fitting formula for the external boundary, ${a_{crit}} (R_H) = c_1 (1 - c_2e_p - c_3e_{sat}$).  The dash symbol (--) denotes when a coefficient is not applicable because the respective eccentricity is fixed at zero.}  
\end{deluxetable}

\begin{deluxetable}{lcccccccc}
\tablecaption{Initial Orbital Elements for Synthetic Lightcurve and RV of Kepler 1625  \label{tab:syn_IC}}
\tablehead{\colhead{} & \colhead{M ($M_\oplus$)} & \colhead{R ($R_\oplus$)} & \colhead{$a$ (AU)} & \colhead{$R_H$ (AU)} & \colhead{$e$} & \colhead{$i$ (deg.)}& \colhead{$\omega$ (deg.)} & \colhead{$\nu$ (deg.)}} 
\startdata
Exoplanet & 1270 & 10 & 0.86348384 & 0.09233077 & 0.01147105 & 0 & 250.31211 & 190.91273\\
Exomoon & 17.91 & 4 & 0.02401178 & 0.00402236 & 0.01159945 & 25.805015 & 264.07880 & 322.26965 \\
Submoon & $10^{-6}$ & 0.0196 & 0.00144350 & 0.00000383 & 0.09984839 & 0.92405241 & 189.40435 & 22.038811 \\
\enddata
\tablecomments{The semimajor axis $a$, Hill Radius $R_H$, eccentricity $e$, inclination $i$, argument of periastron $\omega$, and true anomaly $\nu$ are given relative to the orbital plane of the respective host body, where we set the ascending node $\Omega = 180^\circ$ for all bodies. After a coordinate transformation to a common frame, we rotate the system by $90^\circ$ so that the line-of-sight is along the z-axis.} 
\end{deluxetable}



\end{document}